# Dynamics of Long-Lived Carriers in Molybdenum Carbide Nanosheets

*Xiangyu Zhu,[†,‡] Zhong Wang,[§] Tao Li,[†] Xi Wang,[†,‡] Zheng Zhang,[†,‡] Chunlong Hu,[†,‡] Kaifu Huo,[†] Wenxi Liang*,[†,‡]*

[†]Wuhan National Laboratory for Optoelectronics, Huazhong University of Science and Technology, Wuhan 430074, China

[‡]Advanced Biomedical Imaging Facility, Huazhong University of Science and Technology, Wuhan 430074, China

[§]The Institute of Advanced Light Source Facilities, Shenzhen 518107, China

Corresponding Author

*Email: wxliang@hust.edu.cn

## ABSTRACT

Molybdenum carbide (MoC) is a promising candidate for substituting expensive platinum-group metals in many applications owing to its low cost and excellent properties, therefore a comprehensive understanding of its carrier dynamics is demanded for advancing the synthesis strategy and implementation of MoC. In this work, the carrier relaxation in MoC nanosheets is investigated by combining the femtosecond transient reflection spectroscopy with the first-principles calculations. The measured processes of electron–electron, electron–phonon, and phonon–phonon scattering show significantly longer lifetimes compared to those of other transition metal carbides. Through the analysis of calculated phonon dispersion, the nanosecond carrier lifetime is explained by the restricted phonon decay pathways, which is induced by the large mass difference between Mo and C atoms. The observed slow carrier cooling rate in our study offers a straightforward approach for effectively utilizing hot carriers, which is expected to improve the performance of photothermal and photovoltaic devices.

## TOC GRAPHIC

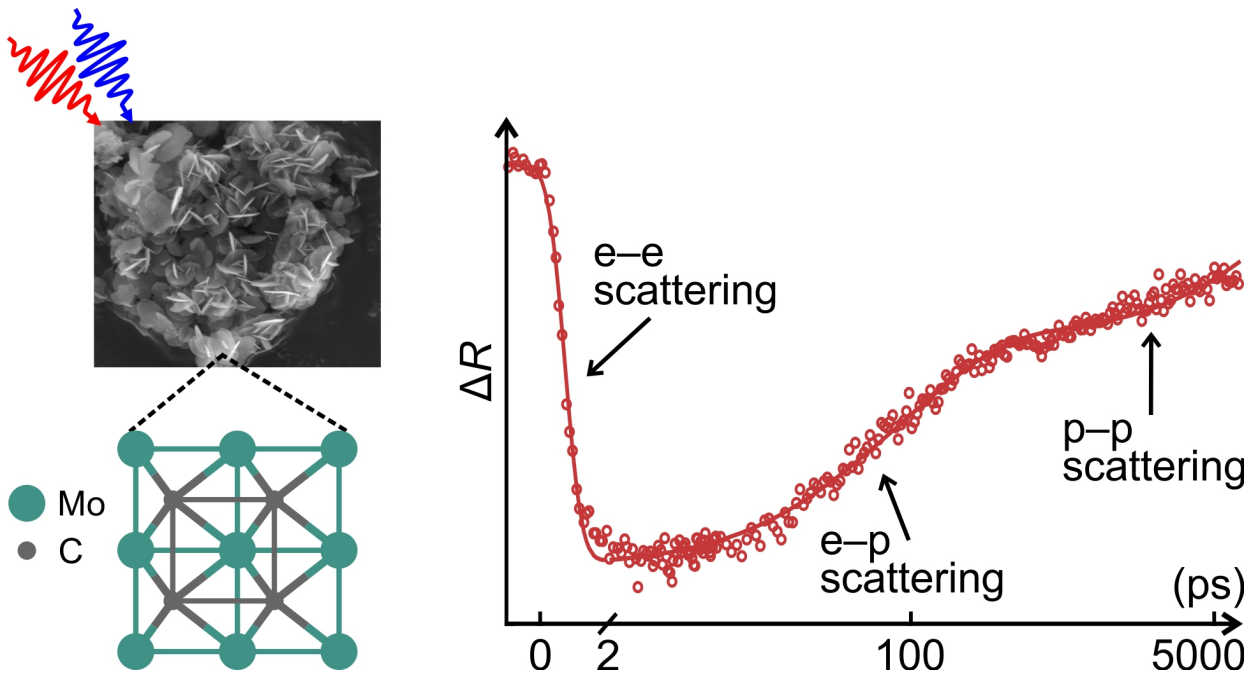

## INTRODUCTION

Transition metal carbides (TMCs), as the emerging members of two-dimensional (2D) material family, hold promising potentials in applications of catalysis, energy storage, flexible electronic devices, and so on,[1,2] owing to their many favored properties including high chemical stability and excellent optoelectronic performance.[3,4] In order to make better use of these materials, progresses have been made in studies on the mechanisms of optical, electrical, and thermal properties of 2D TMCs. For instance, the study of nonlinear broadband absorption of $Ti_3C_2$ nanosheets found that the lifetimes of electron–phonon (e–p) and phonon–phonon (p–p) interactions extended with increasing pump fluences,[5] showing the potential in applications of broadband optical limiter and novel photonic device.[6] Another 2D TMC, $Nb_2C$, which is promising for ultrafast photonic devices, exhibits stronger optical response and higher photothermal conversion efficiency than those of $Ti_3C_2$,[7] while its carrier relaxation time is shorter than that of graphene.[8] So far many 2D TMCs based applications are still in their infancy, as the carrier dynamics of 2D TMCs, which determine their performances, are still under investigation.

In recent years, the development of alternatives to the noble metals in plasmonic nanostructure attracts considerable interests.[9] Molybdenum carbide (MoC) was extensively studied in applications of catalysis and energy storage,[3,10] in view of its advantages of abundance, low cost, and in particular, the similar electronic structure and catalytic properties to those of platinum-group metals.[11] According to the studies of band structure, the work function of MoC is 4.81 eV,[12] and the absence of band gap near Fermi level indicates its metallic properties.[13] Additionally, its steady-state absorption spectra demonstrate the strong absorption that covers a broad wavelength range from visible to near-infrared light (500–2000 nm),[14] marking it an excellent broadband optical material. However, the photophysical properties of MoC remain uncharted, particularly in the hot carrier dynamics and photothermal response on ultrafast timescale. Elucidating these dynamics sets the basis for implementing MoC in applications across photovoltaics to photocatalysis.

In this work, we investigate the carrier dynamics in MoC nanosheets by employing the femtosecond transient reflection (TR) spectroscopy and the first-principles calculations. The bleach signals in TR measurement decay with the processes of e–p and p–p scattering on timescales of hundred-picosecond and a few nanoseconds, respectively. Such long decays suggest a carrier lifetime longer than those of other TMCs, such as $Nb_2C$ and $Ti_3C_2$. The pathways of phonon decay accounting for the long-lived carriers are resolved by the calculation of phonon dispersion, in which we find that the large phonon band gap, introduced by the large mass difference between Mo and C atoms, results in the suppressed phonon decays and the subsequently extended carrier relaxation process.

## METHODS

**Sample preparation and characterization.** The samples of MoC nanosheet were prepared by a $Na_2CO_3$-assisted annealing treatment in $Ar/H_2$ atmosphere, with natural molybdenite (bulk $MoS_2$) as the molybdenum precursor and urea as the carbon precursor. The obtained MoC nanosheets were dispersed in ethanol and transferred onto a glass slide, then dried at 80°C for 5 minutes before measurements.

The morphology of sample was examined using a field emission scanning electron microscope (Nova Nano SEM 450, FEI) and an atomic force microscope (Jupiter XR, Oxford Instruments

Asylum Research). The crystal structure was examined using an X-ray diffractometer (X'Pert3 Powder, PANalytical B.V.). The steady-state absorption spectrum was measured using an ultraviolet-visible-near-infrared spectrophotometer (SolidSpec-3700, Shimadzu).

**Transient reflectivity spectroscopy.** The TR measurements were performed via a femtosecond transient absorption spectrometer (Helios, Ultrafast Systems), working in the reflection mode with a microscope extension. The output of femtosecond laser (800 nm, 40 fs, operating at a repetition rate of 5 kHz, Legend Elite, Coherent), was divided into two beams. One beam with the major part of energy was introduced into an optical parametric amplifier (TOPAS, Light Conversion) to generate pump pulses, which were tunable in 240–2700 nm, then was modulated by a mechanical chopper. The other beam with the small part of energy passed through a delay line to introduce optical path difference, then through a sapphire crystal to generate probe pulses of white-light continuum with wavelength of 420–820 nm. The pump and probe pulses were closely focused on the edge of single MoC flake through a microscope objective in a near-axis configuration, to obtain the clear and strong TR signals. The pump and probe spots on the sample surface were focused with radii of 2 and 1.5 μm, respectively, which were monitored by a charge coupled device (CCD) camera. At the end, the reflectivity signals were recorded by a linear array CCD sensor. The temporal resolution of the whole system was estimated to be ~140 fs by the cross-phase modulation measurement.

**First-principles calculations.** The calculations of geometry optimization of monolayer MoC were performed using the Vienna *ab initio* simulation package (VASP) code[15] with the generalized gradient approximation (GGA) of Perdew–Burke–Ernzerhof (PBE). The projector augmented wave (PAW) method was utilized with a plane-wave basis set,[16] and the convergence criteria were set to $10^{-8}$ eV for energy and $10^{-6}$ eV/Å for force. A kinetic cutoff energy of 400 eV and a Monkhorst–Pack[17] special k-point mesh of 11×11×2 were employed for the calculations. A vacuum layer with thickness of 15 Å was used in the monolayer MoC to avoid interactions between adjacent monolayers. The phonon properties were obtained from the density functional perturbation theory using the Phonopy software package[18] and a 4×4×1 supercell.

## RESULTS AND DISCUSSION

The investigated MoC nanosheets are synthesized through a salt-assisted method (see Methods). Their morphology, crystallization and optical response are examined using scanning electron microscopy (SEM), X-ray diffraction (XRD) and steady-state absorption spectrum, respectively, see Figure 1. The synthesized samples with hexagonal lattice structure tend to agglomerate, as shown in Figure 1a, exhibiting large stacked areas and dispersed small fragments with lateral dimension of ~1 μm. Under XRD examination, strong diffraction peaks that verify the hexagonal phase are observed, as shown in Figure 1b, as well indicating the high crystallinity of samples. The steady-state absorption spectrum shows the strong absorption in a range across visible to near-infrared wavelength, as shown in Figure 1c, indicating the metallic nature of hexagonal MoC.

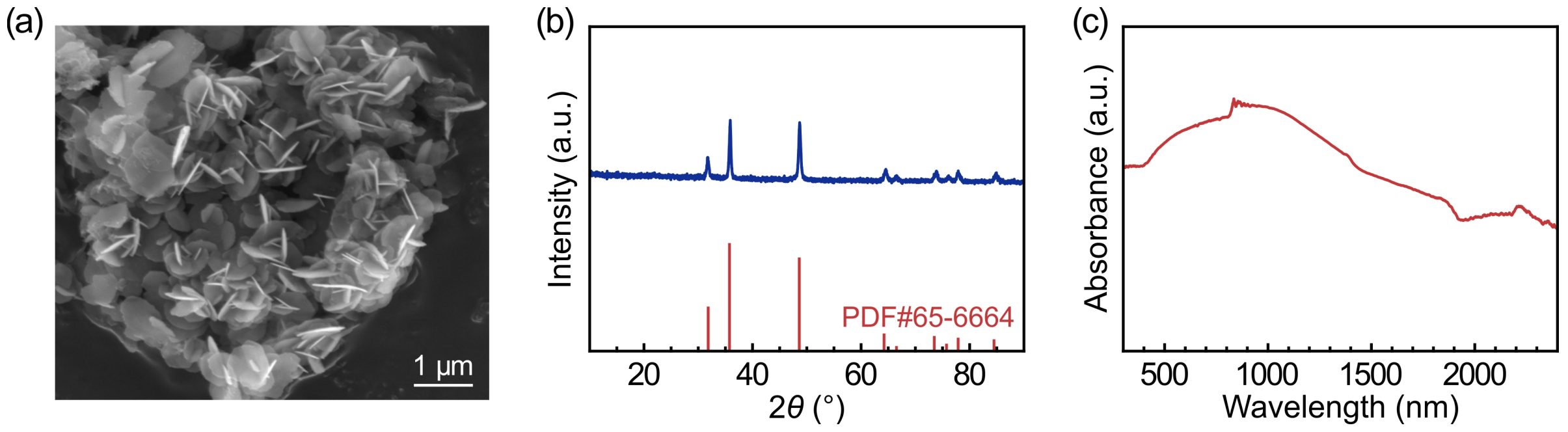

**Figure 1.** Sample characterizations. (a) SEM image, showing densely stacked MoC nanosheets.

The atomic force microscopy measurements show the thickness of single flake less than 2 nm, consisting of 2 layers of MoC (see Figure S1 in Supplementary Materials). (b) XRD pattern compared with the standard PDF data, showing diffraction peaks consistent with the crystal structure of hexagonal MoC. (c) Steady-state absorption spectrum.

We first examine the hot carrier relaxation by the TR measurements with visible supercontinuum white light detection, upon excitation of 400 nm (see Methods). As shown in Figure 2a, the sample appears mainly opaque with shiny reflected light when it is viewed under the microscope. The pseudocolor contour plot of the measured TR spectrum, see Figure 2b, shows a broad bleach signal in visible range dominating the spectrum. Such transient responses match the strong absorption observed in the steady-state absorption spectrum. The bleach signal peaks at ~550 nm, featuring the transient optically-induced broadening (damping) of localized surface plasmon resonance peak,[19] which originates from the electronic transition dynamics. As depicted in Figure 2c, the spectrum profiles at selected delay times show that the bleach amplitude reaches its maximum in a few picoseconds after excitation, then decays in a process that lasts more than 8 ns, showing a long carrier lifetime on the time scale of nanosecond.

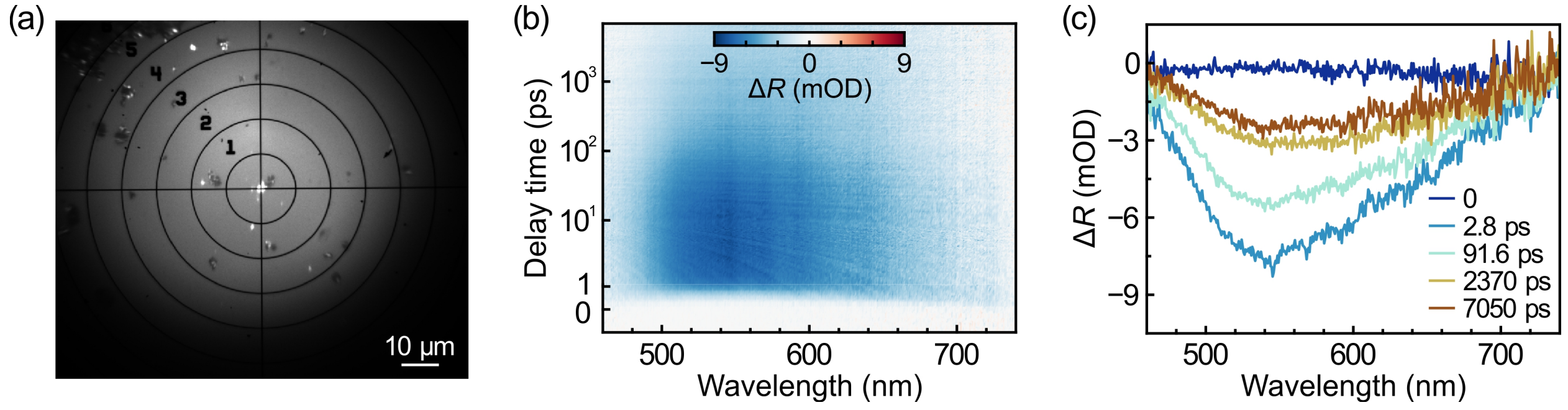

**Figure 2.** TR measurements upon 400 nm excitation. (a) Optical photograph of a single flake of sample (centered white spot) illustrated by excitation pulses, viewed through the microscope. (b)

Pseudocolor contour plot of the measured TR spectrum. (c) Spectrum profiles at selected delay times, extracted from (b).

More details of the bleach signal evolution are examined with various injected carrier densities. Figure 3a depicts the spectra evolution following the increase of excitation power, at the delay time when the bleach amplitude reaches the maximum. A prominent red shift of the bleach peak is observed as the excitation power increases. Similar red shifts were also observed in many metals including Ni, Au, Ag, and Ge,[20–23] and were attributed to the enhanced electron–electron (e–e) scattering and e–p scattering, or the band renormalization effect under high excitation power. Furthermore, the maximum bleach amplitude increases as the excitation power increases, exhibiting an almost linear dependence before reaching the saturation regime, as depicted in the inset of Figure 3b. We extracted the power-dependent kinetics probed at 544 nm, and fit the traces with triexponential functions (see Supplementary Table S1 in Supplementary Materials for tabulated results), finding that the decay processes show a three-stage evolution, as depicted in Figure 3b. Such bleach decays reflect that the photoexcited carriers in MoC relax in three stages with lifetimes of several tens of picoseconds, a few nanoseconds, and several nanoseconds beyond the measured time window, respectively. Compared to the carrier lifetimes of other 2D materials, e.g., a few picoseconds of graphene,[24] a few hundred picoseconds of $MoS_2$[25] and $WS_2$,[26] the carrier lifetime of MoC is significantly longer.

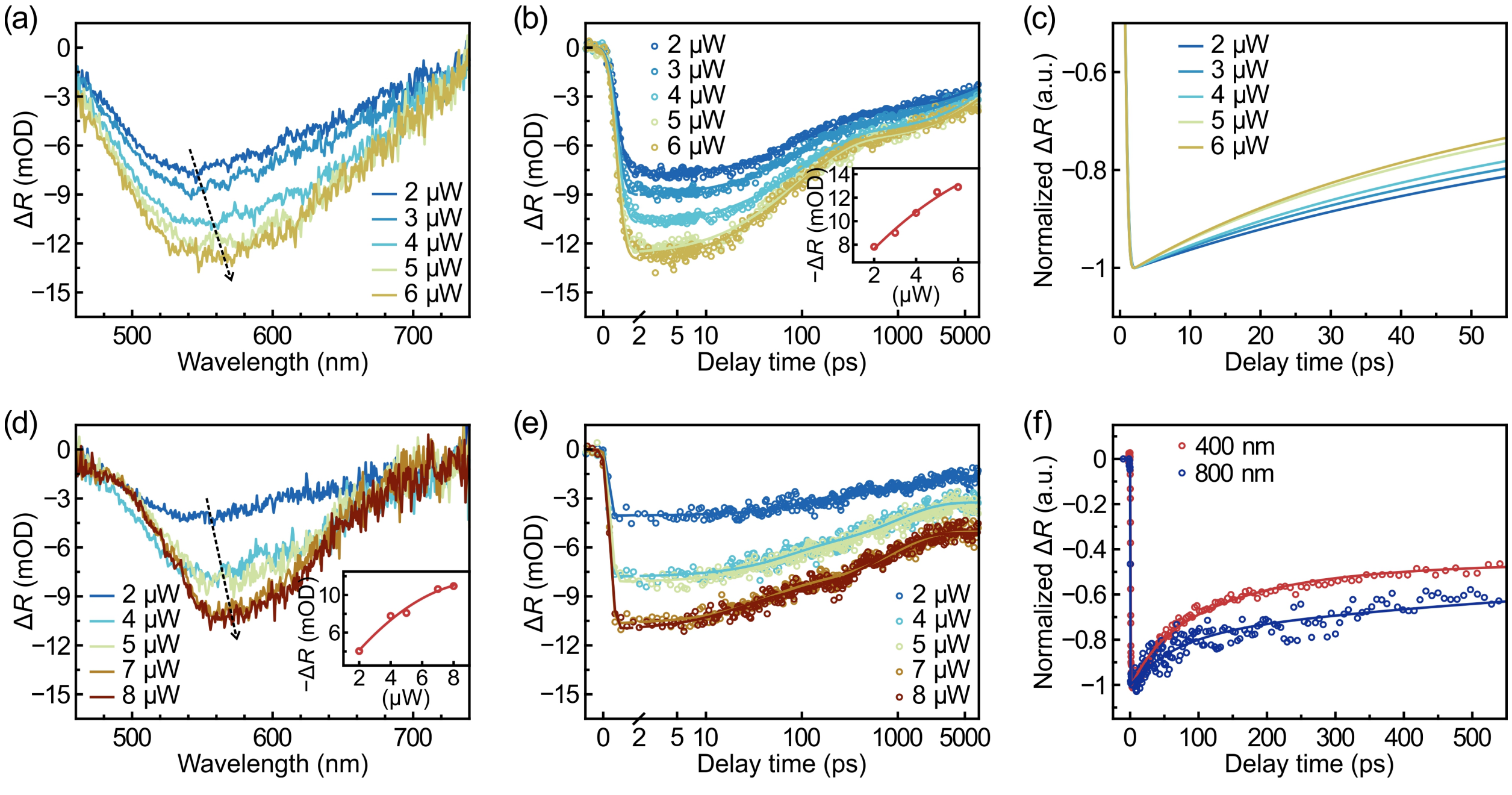

**Figure 3.** Carrier relaxation upon 400 nm (a–c) and 800 nm (d–e) excitations with various powers. (a, d) Evolution of spectrum profiles following the increase of excitation power. Dashed arrow, the guide to the eyes. (b, e) Power-dependent kinetics probed at 544 and 562 nm, respectively. Solid lines, triexponential function fits. (c) Normalized fit curves in (b) at the early several tens of picoseconds. (f) Comparison between the bleach decays excited by 400 nm and 800 nm, with the same excitation power of 4 μW. Solid lines, triexponential function fits. Inset in (b, d), saturating trend of the maximum bleach amplitudes; Solid line, the guide to the eyes.

In metallic materials, there are three common processes in transient responses to photoexcitation,[27] i.e., e–e scattering, e–p scattering, and p–p scattering or heat diffusion, following the sequence of their occurrences. In transient absorption or TR measurements, the e–e scattering is manifested in the rise process of bleach kinetics trace. The characteristic times of bleach rise are on the order of tens to hundreds of femtoseconds for metals and metal compounds, or materials with metallic properties, such as Au, Ag, Al, TiN, and ZrN.[28–30] In our case of MoC, the characteristic rise time ($\tau_0$) varies from ~500 femtoseconds (upon 400 nm excitation, see Supplementary Table S1) to one picosecond (upon 800 nm excitation, see later discussions), indicating a significantly slowed electron equilibration rate, which is a benefit to maintaining the hot electron population for extraction before the excited electrons loss their excess energy.

With increasing excitation power, the bleach decay shows a boosted trend, as demonstrated by the normalized fit curves in Figure 3c. The extracted lifetimes of the first decay stage ($\tau_1$) decrease following the increased injection of carrier density, as summarized in Supplementary Table S1. Such boosted bleach decays are the feature of Auger recombination or e–p scattering process. However, the Auger recombination in 2D materials usually occurs on a faster timescale, e.g., sub-picosecond to a few picoseconds of 2D perovskites[31] and a few picoseconds of $WS_2$,[26] as the high binding energy in 2D materials is associated with stronger electron–hole interactions and an enhanced Auger recombination rate.[31] Therefore, this stage with a time span of several tens of picoseconds in MoC nanosheets is attributed to the e–p scattering. Moreover, the effect of defect trapping is also excluded, as its associated characteristic time increases with increasing excitation power till saturation,[32] in contrast to our observations. In terms of e-p scattering, the measured lifetimes in MoC nanosheets are apparently slower than those in metals or similar materials,[30,33] manifesting a lower e–p scattering rate and subsequently an extended carrier lifetime. Furthermore, the fitted weights of $\tau_1$ are apparently larger than those of $\tau_2$ and $\tau_3$ (see Supplementary Table S1), indicating the dominant role of e–p scattering during the carrier relaxation, i.e., more electrons release their extra energy through the channel of directly coupling to lattice.

Following the e–p scattering, the possible p–p scattering or heat diffusion for the second stage of carrier relaxation also shows decreased lifetime ($\tau_2$, see Supplementary Table S1) following the increase of excitation power. This boosted rate, agreeing well with the feature of accelerated thermalization under the increased injection of carrier density,[34] is attributed to the increased p–p scattering rate due to the raised phonon population,[35] which also introduce the increase of final lattice temperature.

The longest decay lasts several nanoseconds ($\tau_3$, see Supplementary Table S1), representing a carrier lifetime that is significantly longer than those of other TMCs, for instance, a few picoseconds in $Nb_2C$[36] and around ten picoseconds in $Ti_3C_2$.[5] This long lifetime is similar to the long decay times observed in HfN[37] and BAs[38] nanosheets. In HfN and BAs, there is a large phonon band gap between the optical branches (OBs) and the acoustic branches (ABs) due to the large mass difference between Hf (B) and N (As) atoms. As OBs are associated to the displacements of light atoms while ABs to the displacements of heavy atoms,[39] the large phonon band gap limits the pathways for optical phonons (OPs) decaying into acoustic phonons (APs). Consequently, the retarded phonon decay results in re-scattering of extra energy from the OPs back to the electrons, i.e., the occurrence of phonon bottleneck.[38] Given the large mass difference between Mo and C atoms, the conservation of energy and momentum also restricts the phonons directly decay from OBs into ABs, see discussions later. As a result, the relaxation process of photoexcited carriers extends to the time scale of nanosecond.

Then we investigate the impact of initial electron temperature on the relaxation process by comparing the TR responses upon excitations of 400 nm and 800 nm. Figure 3d depicts the spectrum evolution under various excitation powers of 800 nm (at the delay time when the bleach amplitude reaches the maximum), showing the narrowed broadband profiles compared to those upon 400 nm excitation (Figure 3a). A red shift following the increase of excitation power is also observed. Comparing to the case of 400 nm excitation, the maximum bleach amplitude upon 800 nm excitation reach to saturation faster when the excitation power increases, as depicted in the inset of Figure 3d. Moreover, the amplitudes excited by 800 nm with same powers (e.g., 2, 4 and 5 μW) are apparently smaller than those excited by 400 nm. This difference originates from the deferent rises of electron temperature after photoexcitation, as the recorded reflectivity varies linearly with the electron temperature.[30,40] Again, the bleach signal excited by 800 nm shows a three-stage decay (see Figure 3e for the kinetics extracted at 562 nm), with the decreasing lifetimes of the first two stages following the increase of excitation power (see Supplementary Table S2 in Supplementary Materials for tabulated fit results). The decay of these two stages is lower than that upon 400 nm excitation, as compared in Figure 3f. Furthermore, the fitted weights of the first decay are apparently smaller than those upon 400 nm excitation. Both the lowered scattering rate and weight of the first decay indicate that the e–p scattering process is retarded by the low initial electron temperature.

The different carrier relaxations under varied excitation conditions are the straightforward results from carrier-phonon interactions. Compared to absorbing 800 nm photons, the electrons absorbing 400 nm photons populate further away from the Femi level, leading to the higher electron temperature and e–p scattering rate,[41–43] then the higher lattice temperature given the same excitation power. As the high-energy excitation allows for the efficient emissions of OPs during carrier relaxation, the rapid scattering channel with OPs dominates the relaxation pathways[44] and results in the increased scattering rate, while the low-energy excitation restricts the carrier relaxation to the slow scattering channel with APs.[45] Moreover, both the e–p and p–p scattering processes show a trend that their lifetimes gradually stabilize with the increasing excitation power, as depicted in Figure 4. Such a trend may be interpreted by the enhanced hot-phonon bottleneck effect upon high excitation power.[46,47] Under the condition of high injected carrier density, the high phonon emission rate results in a non-equilibrium phonon population,[48,49] which increases the phonon reabsorption rate while reduces the thermalization rate (or the phonon emission rate).[46] Taking this effect into account, we can summarize a scenario that the phonon bottleneck competes with the accelerated electron and lattice thermalizations, leading to the eventually converged lifetimes of carrier relaxation when the injected carrier density increases.

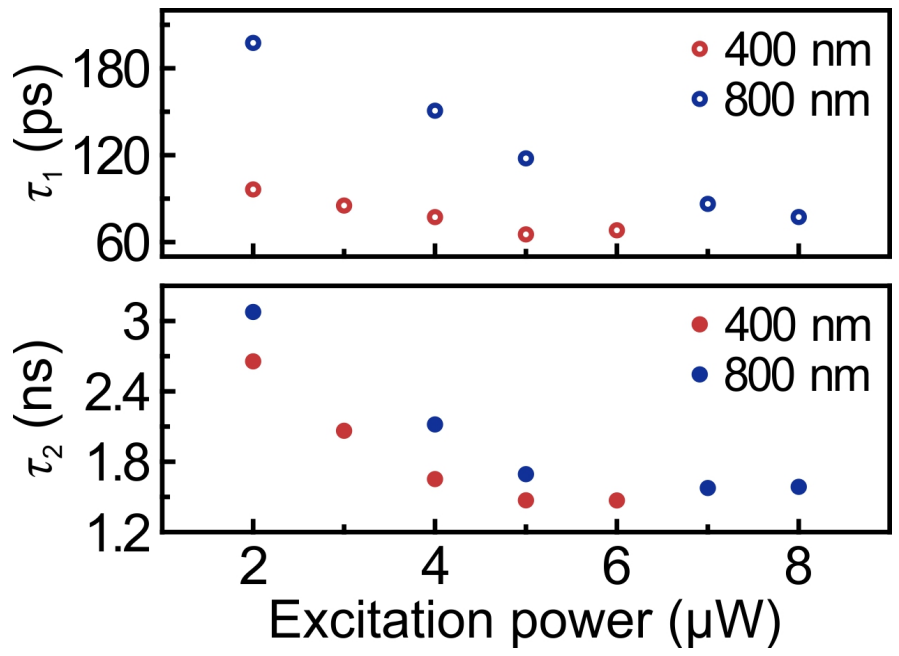

**Figure 4.** Convergence of the e–p and p–p scattering lifetimes following the increase of excitation power.

The pathways of carrier relaxation are further elucidated by the phonon dispersion obtained through the first-principles calculations with a monolayer configuration (see Methods). As depicted in Figure 5, there is a large phonon band gap separating the maximum of ABs and the minimum of OBs. This gap is significantly larger than those of well-studied 2D transition metal sulfides, such as WS and WSe,[51] rendering the maximum of ABs lower than half of the minimum of OBs in MoC. Invoking the cases of HfN and BAs, the large phonon band gap in MoC is probably also induced by the large mass difference between Mo and C atoms. Subsequently, the gap results in the suppressed phonon decay and the occurrence of phonon bottleneck, which in turn hinder the carrier relaxation.

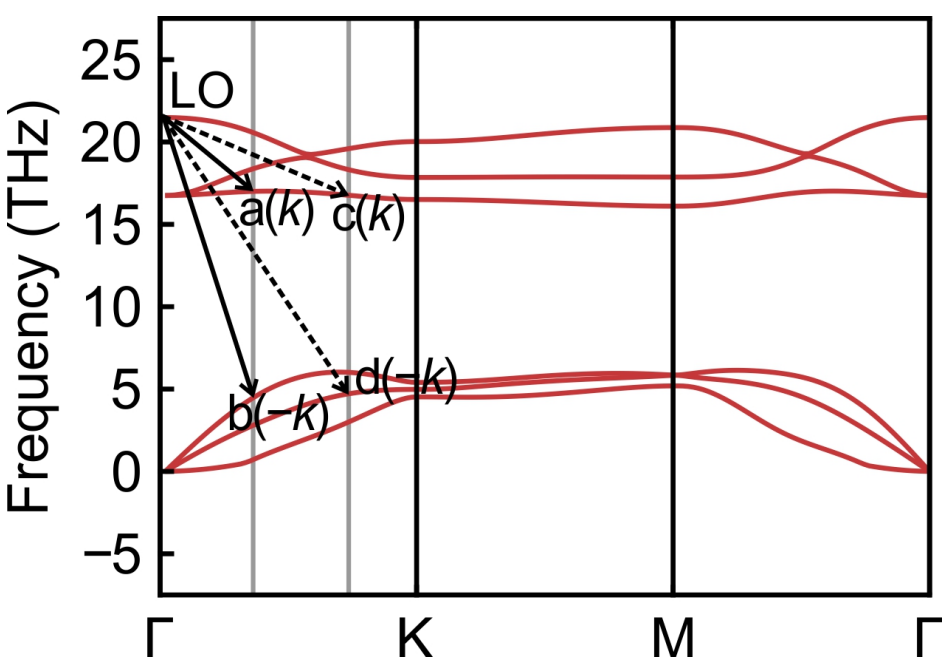

**Figure 5.** Calculated phonon dispersion of monolayer MoC. Both the absence of imaginary frequency mode and the highest vibration frequency of OBs up to 21.9 THz suggest the dynamical stability.[50] Arrows, possible decay pathways for the LO phonons with high frequency, see text.

Specifically, the decay of longitudinal optical (LO) phonons lays significant impacts on the carrier relaxation, as the Fröhlich interaction between carriers and LO phonons is the primary carrier–phonon scattering mechanism in polar materials.[52] Determined by the laws of energy and momentum conservation, the lattice anharmonicity allows for the LO phonons with high frequency decaying into the phonons with lower frequency through various mechanisms, such as the three-phonon process,[53] in which one LO phonon decays into two phonons.[54–56] In general, three-phonon interactions are slower than the processes of LO phonon emission through carrier coupling, and the strength of higher-order processes are even weaker by orders of magnitude.[57,58] Due to the large phonon band gap and the flat OP dispersion (see Figure 5), the channels for generating two low-frequency phonons in the same band, e.g., the Klemens channel,[54,59] are suppressed. Furthermore, the distinct splitting of the LO and transvers optical branches around Γ point facilitates the Ridley–Gupta channel,[55,60] i.e., one LO phonon decays into one OP and one AP. For configuring the possible decay pathways, we draw two lines (in grey) parallel to the y-axis of calculated dispersion, intersecting the OBs (ABs) at points a and c (b and d), see Figure 5. The summed frequencies of a and b (c and d) are equal to the highest frequency at Γ point, ensuring energy conservation. Therefore, the LO phonons with the highest frequency may possibly decay into points a and b (or c and d), with opposite wave vectors governed by momentum conservation, undergoing the slow three-phonon processes. Since both the frequencies of a and c are higher than two-fold the highest frequency of ABs, the phonons decayed into a or c subsequently decay to ABs through four-phonon or higher-order processes, yielding even more significantly slowed relaxation processes.

Thus, the carrier relaxation in MoC nanosheets can be outlined as the following scenario. Initially, the photoexcited carriers transit into the high-energy states and rapidly achieve a quasi-equilibrium distribution through e–e scattering. Then the hot carriers transfer their energy to lattice through e–p coupling in tens to hundreds of picoseconds, basing on the temperature of excited carriers. Subsequently, the phonons with high frequency decay into the low-energy phonons primarily through the slow channels of multiple-phonon interactions on the time scale of nanosecond, at the end resulting in the long-lived carriers with lifetime beyond the measured time window of several nanoseconds.

**CONCLUSIONS**

In summary, the relaxation of photoexcited carriers in MoC nanosheets is elucidated by the combination of femtosecond TR measurements and first-principles calculations. The revealed slow carrier relaxation originates from the suppressed e–p and p–p scattering due to the large phonon band gap, possibly accompanied with the phonon bottleneck effect. The long-lived carriers make possible the implementation of MoC nanosheets in photothermal and photovoltaics applications, particularly serving as an absorbing material for the potential hot-carrier solar cells, in which the slow cooling of hot carriers is a key factor for improving carrier extraction.

## ASSOCIATED CONTENT

### Supporting Information

Additional Figure S1, Tables S1 and S2 (PDF).

## AUTHOR INFORMATION

### Corresponding Author

*Email: wxliang@hust.edu.cn

### Author Contributions

X.Z. and W.L. conceived of the project. X.Z. performed the measurements with supports from X.W., Z.Z., and C.H. T.L. synthesized the MoC nanosheets under the supervision from K.H. Z.W. performed the first-principles calculations. X.Z. and W.L. analyzed the data with discussions with all authors. X.Z. and W.L. wrote the paper with contributions from all authors. W. L. supervised the project.

## Notes

The authors declare no competing financial interest.

## ACKNOWLEDGMENTS

X.Z., C.H. and W.L. thank the National Key R&D Program of China (2025YFF0514900) for support. T.L. and K.H. thank the National Natural Science Foundation of China (52572225 and U2004210) for support. We thank the Analytical and Testing Center in Huazhong University of Science and Technology for support.

*Supplementary Materials for*

# Dynamics of Long-Lived Carriers in Molybdenum Carbide Nanosheets

*Xiangyu Zhu,[†,‡] Zhong Wang[§] Tao Li,[†] Xi Wang,[†,‡] Zheng Zhang,[†,‡] Chunlong Hu,[†,‡] Kaifu Huo,[†]*

*Wenxi Liang*,[†,‡]*

[†]Wuhan National Laboratory for Optoelectronics, Huazhong University of Science and Technology, Wuhan 430074, China

[‡]Advanced Biomedical Imaging Facility, Huazhong University of Science and Technology, Wuhan 430074, China

[§]The Institute of Advanced Light Source Facilities, Shenzhen 518107, China

Corresponding Author

*Email: wxliang@hust.edu.cn

## Supplementary Figures and Tables

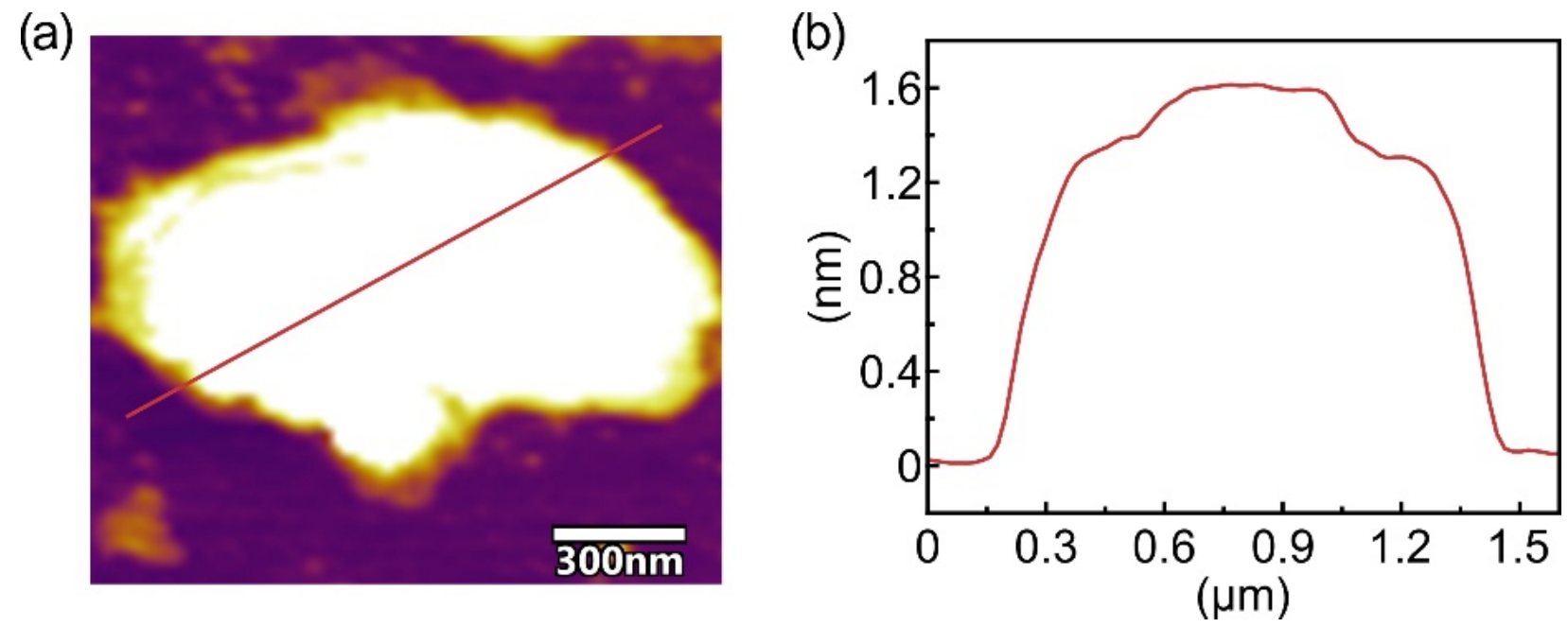

**Figure S1.** Sample characterization by atomic force microscopy (AFM). (a) AFM image of a single flake of MoC. (b) Thickness extracted from the region marked with the red line in (a), showing a uniform thickness of ~1.5 nm.

**Table S1.** Fit results of the kinetics probed at 544 nm upon 400 nm excitation.

| Power (μW) | $A_0$ | $\tau_0$ (ps) | $A_1$ | $\tau_1$ (ps) | $A_2$ | $\tau_2$ (ps) | $A_3$ | $\tau_3$ (ps) |
|---|---|---|---|---|---|---|---|---|
| 2 | 100% | 0.5080 | 44.28% | 96.35 | 25.92% | 2656 | 29.80% | > 8000 |
| 3 | 100% | 0.5007 | 42.86% | 85.22 | 25.51% | 2064 | 31.63% | > 8000 |
| 4 | 100% | 0.4977 | 44.08% | 77.25 | 25.10% | 1652 | 30.82% | > 8000 |
| 5 | 100% | 0.4819 | 45.81% | 65.33 | 25.15% | 1471 | 29.04% | > 8000 |
| 6 | 100% | 0.4793 | 46.08% | 68.15 | 25.49% | 1470 | 28.43% | > 8000 |

**Table S2.** Fit results of the kinetics probed at 562 nm upon 800 nm excitation.

| Power (μW) | $A_0$ | $\tau_0$ (ps) | $A_1$ | $\tau_1$ (ps) | $A_2$ | $\tau_2$ (ps) | $A_3$ | $\tau_3$ (ps) |
|---|---|---|---|---|---|---|---|---|
| 2 | 100% | 1.316 | 34.56% | 197.5 | 30.32% | 3077 | 35.12% | > 8000 |
| 4 | 100% | 1.113 | 33.69% | 150.7 | 29.30% | 2118 | 37.01% | > 8000 |
| 5 | 100% | 1.025 | 28.53% | 117.8 | 30.81% | 1694 | 40.66% | > 8000 |
| 7 | 100% | 0.8751 | 26.22% | 86.38 | 31.13% | 1576 | 42.65% | > 8000 |